\documentclass[hyper,letterpaper,notoc]{JHEP3}

\usepackage{epsfig}
\usepackage{amsbsy}
\usepackage{varioref}
\usepackage{pifont}

\newcommand{\dfrac}[2]{\frac{\strut \displaystyle{#1}}{\strut \displaystyle{#2}}}

\newcommand{\lessim}{\mathop{<}\limits_{\displaystyle{\sim}}}

\def\fsl#1{\setbox0=\hbox{$#1$}           
   \dimen0=\wd0                                 
   \setbox1=\hbox{/} \dimen1=\wd1               
   \ifdim\dimen0>\dimen1                        
      \rlap{\hbox to \dimen0{\hfil/\hfil}}      
      #1                                        
   \else                                        
      \rlap{\hbox to \dimen1{\hfil$#1$\hfil}}   
      /                                         
   \fi}                                         %

\title{Deconstructed Higgsless Models with \\
One-Site Delocalization
}

\author{R. Sekhar Chivukula and Elizabeth H. Simmons\\
Department of Physics and Astronomy, Michigan State University\\
East Lansing, MI 48824, USA\\
	E-mail: \email{sekhar@msu.edu, esimmons@msu.edu}}
	
\author{Hong-Jian He\\
Department of Physics, University of Texas\\
Austin, TX 78712, USA\\
	E-mail: \email{hjhe@physics.utexas.edu}}

\author{Masafumi Kurachi\\
Department of Physics, Nagoya University\\
Nagoya 464-8602, Japan\\
	E-mail: \email{kurachi@eken.phys.nagoya-u.ac.jp}}

\author{Masaharu Tanabashi\\
Department of Physics, Tohoku University\\
Sendai 980-8578, Japan\\
	E-mail: \email{tanabash@tuhep.phys.tohoku.ac.jp}}

\abstract{
In this note we calculate the form of electroweak corrections
in deconstructed Higgsless models
for the case of a fermion whose weak properties arise from two adjacent $SU(2)$
groups on the deconstructed lattice. 
We show that, as recently proposed in the continuum, it is possible for the value
of the electroweak parameter $\alpha S$ to be  small in such  a model. In addition,  
by working in the deconstructed
limit, we may directly evaluate the size of off-$Z$-pole electroweak corrections
arising from the exchange of Kaluza-Klein modes; this has not been studied in the continuum. The
size of these corrections is summarized by the electroweak parameter
$\alpha \delta$.  In one-site delocalized Higgsless models with small values of
$\alpha S$, we show that the amount of
delocalization is bounded  from above, and must be less than 25\% at 95\% CL.
We discuss the relation of these calculations to our previous calculations in
deconstructed Higgsless models, and to models of extended technicolor.
We present numerical results for a four-site model, illustrating our
analytic calculations.
}

\keywords{Dimensional Deconstruction, Electroweak Symmetry Breaking, Higgsless Theories, Delocalization}

\preprint{MSUHEP-050217\\
TU-739\\
DPNU-05-02}

\begin{document}

\section{Introduction}

``Higgsless'' models \cite{Csaki:2003dt} have emerged as an intriguing direction for research into the origin of electroweak symmetry breaking.  In these models, which are based on five-dimensional gauge theories compactified on an interval, unitarization of the electroweak bosons' self-interactions occurs 
through the exchange of a tower of massive vector bosons 
\cite{SekharChivukula:2001hz,Chivukula:2002ej,Chivukula:2003kq,He:2004zr}, 
rather than the exchange of a scalar Higgs boson \cite{Higgs:1964ia}.  

We have recently analyzed the electroweak corrections in a large class of Higgsless models in which the fermions are localized within the extra  dimension \cite{SekharChivukula:2004mu,Chivukula:2004pk,Chivukula:2004af}.  Specifically, we studied all Higgsless models which can be deconstructed \cite{Arkani-Hamed:2001ca,Hill:2000mu} to a  chain of $SU(2)$ gauge groups adjacent to a chain of $U(1)$ gauge groups,  with the fermions coupled to any single $SU(2)$ group and to any single $U(1)$ group along the chain.  Our use of deconstruction allowed us to relate the size of corrections to electroweak processes directly to the spectrum of vector bosons (``Kaluza-Klein modes") which, in Higgsless models,
is constrained by unitarity.   Our results apply for arbitrary background 5-D geometry,
spatially dependent gauge-couplings, and brane kinetic energy terms. 

We found \cite{SekharChivukula:2004mu} that Higgsless models with localized fermions which do not have extra light vector bosons (with masses of order the $W$ and $Z$ masses) cannot simultaneously satisfy the constraints of precision electroweak data and unitarity bounds.
In particular, we found that unitarity constrains the electroweak parameter
$\hat{S}$ as follows
\begin{equation}
\hat{S} ={1\over 4s^2}\left(\alpha S + 4 c^2 (\Delta \rho - \alpha T) + {\alpha \delta \over c^2}\right) 
\ge M^2_W \Sigma_r \ge {M^2_W\over 8 \pi v^2} \simeq 4 \times 10^{-3}~.
\label{eq:final}
\end{equation}
This large a value is disfavored by  precision electroweak data \cite{Barbieri:2004qk}. 

Although we framed those results in terms of their application to continuum Higgsless
5-D models, they also apply far from the continuum limit
when only a few extra vector bosons are present. As such, these
results form a generalization of phenomenological analyses \cite{Chivukula:2003wj} of models of extended  electroweak gauge symmetries \cite{Casalbuoni:1985kq,Casalbuoni:1996qt,Casalbuoni:2004id}  motivated by models of hidden local symmetry \cite{Bando:1985ej,Bando:1985rf,Bando:1988ym,Bando:1988br,Harada:2003jx}.
Our previous results are complementary to, and more general than,
 the analyses of the phenomenology of these models in the continuum
\cite{Csaki:2003zu,Nomura:2003du,Barbieri:2003pr,Davoudiasl:2003me,Foadi:2003xa,Burdman:2003ya,Davoudiasl:2004pw,Barbieri:2004qk,Hewett:2004dv}.   They also apply independent of the form of the high-energy completion of the Higgsless theory; the potentially large higher-order corrections expected to be present in QCD-like completions have been discussed in \cite{Perelstein:2004sc}.

It has been proposed  \cite{Cacciapaglia:2004rb,Foadi:2004ps} that  the size of
corrections to electroweak processes may be reduced by allowing for delocalized
fermions.  We now investigate this possibility in the context of deconstruction.   This paper will focus on the effects of adding fermion delocalization to the deconstructed models which our earlier work identified as having the greatest phenomenological promise (i.e., those in which the electroweak corrections are smallest).  These are models (designated ``Case I'')  in which the fermions' hypercharge interactions are with the $U(1)$ group at the interface between the $SU(2)$ and $U(1)$ groups, and in which the gauge couplings of that $U(1)$ group and of the $SU(2)$ group farthest from the interface are small.  For simplicity, we will assume, in this paper, that the $U(1)$ group adjacent to the interface is the only hypercharge group in the model; this corresponds to taking the $M=0$ limit of the more general models we studied previously \cite{SekharChivukula:2004mu}. We also assume that the
fermions derive their weak properties from two adjacent $SU(2)$
groups in the deconstructed model -- {\it i.e.}, we consider ``one-site'' delocalization.

We have found several relationships between delocalization and electroweak corrections, some confirming what has been found in the continuum and others entirely new.
We confirm that it is possible for the value of the electroweak parameter $\alpha S$ to be  small in models including fermion delocalization; this has been shown already in the continuum \cite{Cacciapaglia:2004rb,Foadi:2004ps}.  
By working in the deconstructed
limit, we may directly evaluate the size of electroweak corrections away from the $Z$-peak which arise from the exchange of Kaluza-Klein modes; this has not previously been examined in the continuum. The
size of these corrections is summarized by the electroweak parameter
$\alpha \delta$ \cite{Chivukula:2004af,Barbieri:2004qk}, which describes flavor-universal non-oblique corrections. 
In one-site delocalized Higgsless models with small values of
$\alpha S$, we show that the amount of
delocalization is bounded from above by a combination of experimental limits on $\alpha\delta$ and the need to ensure that the scattering of longitudinal $W$ bosons is properly unitarized.  At 95\% CL, the amount of delocalization cannot exceed 25\%. 
We discuss the relation of these calculations to our previous calculations in
deconstructed Higgsless models, and to models of extended technicolor. We defer
to a subsequent work \cite{SekharChivukula:2005xm,newwork} the study of multi-site or 
flavor non-universal delocalization, and the generation of fermion masses.\footnote{Flavor 
non-universal interactions will be required in order to generate the diverse
fermion masses. Depending on how this is done, these new interactions may lead
to additional flavor non-universal electroweak corrections \protect\cite{Cacciapaglia:2004rb}.}

In the next two sections we discuss the structure of the gauge and fermion sectors
of the model, and specify the limit in which we analytically compute the size
of corrections to electroweak interactions. In sections 4, 5, and 6, we compute the electroweak
parameters $\alpha S$, $\alpha T$, and $\alpha \delta$, respectively.\footnote{The fourth
such parameter, $\Delta \rho$, is identically equal to 0
in Case I models \protect\cite{SekharChivukula:2004mu}.} In section 7 we discuss
the interpretation of these models, and discuss how such effects can arise in technicolor
theories. In section 8 we present numerical results for a four-site model, illustrating our
analytic calculations and demonstrating explicitly that some models with vanishing $\alpha S$
can have relatively large values of $\alpha \delta$. Section 9 discusses our conclusions and
outlines future work.


\section{Review of the Gauge Sector of the Model}

\EPSFIGURE[h]{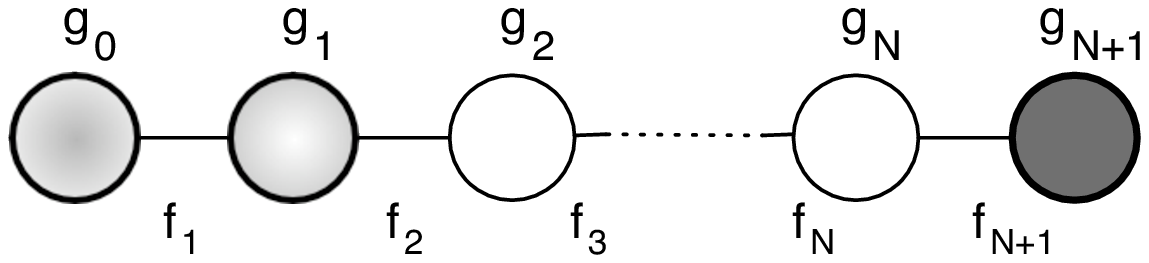,width=0.6\textwidth}
{Moose diagram of the model analyzed in this paper. Sites $0$ to $N$ are
$SU(2)$ gauge groups, site $N+1$ is a $U(1)$ gauge group. The fermions are one-site-delocalized
in the sense that the $SU(2)$ couplings of the fermions
arise from the gauge groups at sites 0 and 1. The $U(1)$ coupling comes from the
gauge group at site $N+1$. \label{fig:tone}}

We study a deconstructed Higgsless model,
as shown diagrammatically (using ``moose notation'' \cite{Georgi:1985hf}) in Figure~ \ref{fig:tone}.   The model incorporates an
$SU(2)^{N+1} \times U(1)$ gauge group, and $N+1$ 
nonlinear $(SU(2)\times SU(2))/SU(2)$ sigma models in which the global symmetry groups 
in adjacent sigma models are identified with the corresponding factors of the gauge group.
The Lagrangian for this model to leading order is given by
\begin{equation}
  {\cal L}_2 =
  \frac{1}{4} \sum_{j=1}^{N+1} f_j^2 \mbox{tr}\left(
    (D_\mu U_j)^\dagger (D^\mu U_j) \right)
  - \sum_{j=0}^{N+1} \dfrac{1}{2g_j^2} \mbox{tr}\left(
    F^j_{\mu\nu} F^{j\mu\nu}
    \right),
\label{lagrangian}
\end{equation}
with
\begin{equation}
  D_\mu U_j = \partial_\mu U_j - i A^{j-1}_\mu U_j 
                               + i U_j A^{j}_\mu,
\end{equation}
where all  gauge fields $A^j_\mu$ $(j=0,1,2,\cdots, N+1)$ are dynamical. The first
$N+1$ gauge fields ($j=0,1,\ldots, N$) correspond to $SU(2)$ gauge groups; the last gauge
field ($j= N+1$) corresponds to the  $U(1)$ gauge group.  The symmetry breaking between
the $A^{N}_\mu$ and $A^{N+1}_\mu$ follows an $SU(2)_L \times SU(2)_R/SU(2)_V$ symmetry
breaking pattern with the $U(1)$ embedded as the $T_3$-generator of $SU(2)_R$.   
Our analysis proceeds for arbitrary values of the gauge couplings and $f$-constants, 
and therefore allows for arbitrary background 5-D geometry,
spatially dependent gauge-couplings, and brane kinetic energy terms for the gauge-bosons.

All four-fermion processes, including those relevant for the electroweak phenomenology of our model, depend on the neutral and charged gauge field
propagator matrices
\begin{equation}
D^Z(Q^2) \equiv \left[ Q^2\, {\cal I} + M^2_{Z}\right]^{-1}~, \ \ \ \ \ \ \ \ 
{D}^W(Q^2) \equiv \left[ Q^2\, {\cal I} + {M}^2_{W}\right]^{-1}~.
\end{equation}
Here, $M_{Z}^2$ and ${M}_W^2$ are, respectively, the mass-squared matrices for the neutral and charged gauge bosons and ${\cal I}$ is the identity matrix.  Consistent with \cite{Chivukula:2004pk}, $Q^2 \equiv -q^2$ refers to the
euclidean momentum. 

The neutral vector meson mass-squared matrix is of dimension $(N+2) \times (N+2)$ 
\begin{equation}
{\tiny
M_{Z}^2 = {1\over 4}
\left(
\begin{array}{c|c|c|c|c}
g^2_0 f^2_1& -g_0 g_1 f^2_1 & & \\ \hline
-g_0 g_1 f^2_1  & g^2_1(f^2_1+f^2_2) & & \\ \hline
 & \ddots & \ddots & \ddots \\ \hline
 & & -g^{}_{N-1} g^{}_{N} f^2_{N} & g^2_{N}(f^2_{N} + f^2_{N+1}) & -g^{}_{N} g^{}_{N+1} f^2_{N+1} \\ \hline
  & & & -g^{}_{N} g^{}_{N+1} f^2_{N+1} & g^2_{N+1} f^2_{N+1}
 \end{array}
\right).
}
\label{eq:neutralmatrix}
\end{equation}
and the charged current vector bosons' mass-squared matrix is the left-upper $(N+1)  \times (N+1) $ dimensional block of the neutral current $M_{Z}^2$ matrix.
The neutral mass matrix (\ref{eq:neutralmatrix}) 
is of a familiar form that has a vanishing determinant, due to a zero eigenvalue.
Physically, this corresponds to a massless neutral gauge field -- the photon.
The non-zero eigenvalues of $M^2_{Z}$
are labeled by ${\mathsf m}^2_{Zz}$ ($z=0,1,2,\cdots, N$), while
those of ${M}^2_W$ are labeled by ${\mathsf m}^2_{Ww}$ ($w=0, 1,2,\cdots, N$). 
The lowest massive eigenstates corresponding to eigenvalues ${\mathsf m}^2_{Z0}$ and 
${\mathsf m}^2_{W0}$ are, respectively, identified as the usual $Z$ and $W$ bosons.
We will generally refer to  these last eigenvalues by their conventional symbols $M^2_Z,\, M^2_{W}$; the distinction between these and the corresponding mass matrices should be clear from
context.

Generalizing the usual mathematical
notation for ``open'' and ``closed'' intervals, we may denote \cite{SekharChivukula:2004mu} the
neutral-boson mass matrix $M^2_Z$ as $M^2_{[0,N+1]}$ --- {\it i.e.},
it is the mass matrix for the entire moose running from site $0$ to site $N+1$ including
the gauge couplings of both endpoint groups. Analogously, the charged-boson mass matrix $M^2_W$ is
$M^2_{[0,N+1)}$ --- it is the mass matrix for the moose running from site $0$ to link
$N+1$, but not including the gauge couping at site $N+1$.
This notation will be useful in thinking about the properties of sub-matrices $M^2_{[0,i)}$ of the full gauge-boson mass matrices that 
arise in our discussion of fermion delocalization, and also the corresponding eigenvalues 
${\mathsf m}^2_{i\,\hat{i}}\ (\hat{i}=1,2,\ldots, i)$.  We will denote the lightest such eigenvalue ${\mathsf m}^2_{i1}$ by the symbol $M^2_i$.

\section{A Moose with Delocalized Fermions}

Consider the simplest deconstructed Higgsless model with one-site delocalized fermions,
as shown in Figure~ \ref{fig:tone}. We take the fermion coupings in this model to 
be
\begin{equation}
{\cal L}_f = \vec{J}^\mu_L \cdot \left({x_0} A^0_\mu + x_1  A^1_\mu\right)
+ J^\mu_Y A^{N+1}_\mu~,
\label{eq:fcoupling}
\end{equation}
where $x_0 + x_1 = 1$ and the fermions are ``delocalized'' in the sense that their $SU(2)$-couplings arise from
both sites 0 and 1. 
Note that the fermion coupings are flavor universal.
This expression is not separately gauge invariant under $SU(2)_0$ and
$SU(2)_1$. Rather, as discussed further in section  \ref{sec:beyond},  eqn. (\ref{eq:fcoupling})
should be viewed as the form of the fermion coupling in ``unitary'' gauge.
Here $\vec{J}^\mu_L$ denotes the isotriplet of left-handed 
weak fermion currents, and $J^\mu_Y$ is the fermion hypercharge current. In the notation
of reference \cite{SekharChivukula:2004mu}, where the fermion coupled to the
$SU(2)$ group at site $p$, the current model is an admixture of $p=0$ and $p=1$. As we
will see, our results for the electroweak parameters in this model are themselves an admixture
of the results we would obtain in the two models.\footnote{The idea of a delocalized model as an admixture of localized-fermion models corresponding to different values of $p$ generalizes readily to multi-site delocalization.  The generalization of the form of equations (3.1) - (3.7) is obvious; the implications will be discussed in a forthcoming paper. \cite{newwork}}

Because  fermions are charged under SU(2) gauge groups at sites 0 and 1, as well as under the single U(1) group at the $N+1$ site, neutral current four-fermion processes may be derived from the Lagrangian
\begin{eqnarray}
{\cal L}_{nc}  & = & - {1\over 2} \left[ \sum_{i,j = 0}^1 x_i x_j g_i g_j \, D^Z_{i,j}(Q^2) \right] 
J^\mu_3 J_{3\mu} - \left[ \sum_{i=0}^1 x_i g_i g_{N+1}\, D^Z_{i,N+1}(Q^2)\right]  J^\mu_3 J_{Y\mu} \nonumber \\
&  &  \quad - {1\over 2} \left[ g^2_{N+1}\, D^Z_{N+1,N+1}(Q^2)\right]  J^\mu_Y J_{Y\mu}~,
\label{nclagrangian}
\end{eqnarray}
and charged-current process from
\begin{equation}
{\cal L}_{cc} = - {1\over 2} \left[ \sum_{i,j=0}^{1} x_i x_j g_i g_j \, {D}^W_{i,j}(Q^2)\right]
 J^\mu_+ J_{-\mu}~.
\label{cclagrangian}
\end{equation}
where $D_{i,j}$ is the $(i,j)$ element of the appropriate gauge field propagator matrix.
We can define correlation functions between fermion currents at given sites as
\begin{equation}
[G_{NC}(Q^2)]_{i,j} = g_i g_j D^Z_{i,j} (Q^2)\qquad \qquad [G_{CC}(Q^2)]_{i,j} = g_i g_j D^W_{i,j}(Q^2)\ .
\end{equation}
The full correlation function for the fermion currents $J^\mu_3$ and $J^\mu_Y$ is then 
\begin{equation}
[G_{\rm NC}(Q^2)]_{WY}= \sum_{i=0}^1 x_i  [G_{\rm NC}(Q^2)]_{i,N+1} ~,
\label{eq:gncwyii}
\end{equation}
where we have used eqn. (\ref{eq:fcoupling}) to include the appropriate contribution from each site to which fermions couple. Likewise, the full correlation function for weak currents is
\begin{equation}
[G_{\rm NC, CC}]_{WW} = \sum_{i,j=0}^1 x_i x_j [G_{\rm NC, CC}]_{i,j}~.
\label{eq:usefulsum}
\end{equation}
The hypercharge correlation function $[G_{NC}(Q^2)]_{YY} = [G_{NC}(Q^2)]_{N+1,N+1}$ depends only on the single site with a $U(1)$ gauge group.

The correlation functions may be written in a spectral decomposition in terms of the mass eigenstates as follows:
\begin{equation}
[G_{\rm NC}(Q^2)]_{YY} =  \dfrac{[\xi_\gamma]_{YY}}{Q^2} 
+\dfrac{[\xi_Z]_{YY}}{Q^2 + M_Z^2} 
+\sum_{z=1}^{N} \dfrac{[\xi_{Zz}]_{YY}}{Q^2 + {\mathsf m}_{Zz}^2},
\label{eq:NC_YY} 
\end{equation}
\begin{equation} 
  [G_{\rm NC}(Q^2)]_{WY} = \dfrac{[\xi_\gamma]_{WY}}{Q^2} 
  +\dfrac{[\xi_Z]_{WY}}{Q^2 + M_Z^2} 
  +\sum_{z=1}^{N} \dfrac{[\xi_{Zz}]_{WY}}{Q^2 + {\mathsf m}_{Zz}^2},
\label{eq:NC_WY}  
\end{equation}
\begin{equation}
  [G_{\rm NC}(Q^2)]_{WW} =  \dfrac{[\xi_\gamma]_{WW}}{Q^2} 
 +\dfrac{[\xi_Z]_{WW}}{Q^2 + M_Z^2} 
  +\sum_{z=1}^{N} \dfrac{[\xi_{Zz}]_{WW}}{Q^2 + {\mathsf m}_{Zz}^2},
\label{eq:NC_WW}  
\end{equation}
\begin{equation}
  [G_{\rm CC}(Q^2)]_{WW} =  \dfrac{[\xi_W]_{WW}}{Q^2 + M_W^2} 
  +\sum_{w=1}^{N} \dfrac{[\xi_{Ww}]_{WW}} {Q^2 + {\mathsf m}_{Ww}^2},
\label{eq:CC_WW}  
\end{equation}
All poles should be simple (i.e. there should be no degenerate mass eigenvalues) because, in the continuum limit, we are analyzing a self-adjoint operator on a finite interval.  
Since the neutral bosons couple to only two currents, $J^\mu_3$ and $J^\mu_Y$, 
the three sets of residues in equations (\ref{eq:NC_YY})--(\ref{eq:NC_WY}) must be related. 
Specifically, they satisfy the $N+1$ consistency conditions,
\begin{equation}
  [\xi_Z]_{WW} [\xi_Z]_{YY}
  = \left([\xi_Z]_{WY}\right)^2, \qquad
  [\xi_{Z{z}}]_{WW} [\xi_{Z{z}}]_{YY}
  = \left([\xi_{Z{z}}]_{WY}\right)^2 .
\label{consistency}
\end{equation}
In the case of the photon, charge universality further implies
\begin{equation}
  e^2 = [\xi_\gamma]_{WW} = [\xi_\gamma]_{WY} = [\xi_\gamma]_{YY}.
\label{eq:universality}
\end{equation}

\subsection{Notation}

We will  find it useful to define the following sums over heavy eigenvalues for phenomenological discussions:
\begin{equation}
\Sigma_{(i,j)} \equiv {\rm Tr}\, M^{-2}_{(i,j)}
\label{eq:firs-sig-def}
\end{equation}
with similar definitions for $\Sigma_{[i,j)}$ and so on.   In particular,
\begin{equation}
\Sigma_{Z} \equiv \sum_{z=1}^N  {1\over {\mathsf m}^2_{Zz}}\ ,\ \ \ \ \ 
\Sigma_{W} \equiv  \sum_{w=1}^N {1\over {\mathsf m}^2_{Ww}}\, ;
\label{eq:sig-def}
\end{equation}
that is, $\Sigma_Z$ and $\Sigma_W$ are the sums over inverse-square masses of the higher neutral- and charged-current KK modes of the full model.
Furthermore, by explicit calculation one finds (see Appendix B of Ref. \cite{SekharChivukula:2004mu})
\begin{equation}
\Sigma_{(0,N+1)} = \sum_{i=1}^N {4 F^2 \over g^2_i F^2_i \tilde{F}^2_i}~,
\label{eq:sigma}
\end{equation}
where
\begin{equation}
{1\over F^2_i} = \sum_{j=i+1}^{N+1} {1\over f^2_j}~,\qquad
{1\over \tilde{F}^2_i} = \sum_{j=1}^i {1\over f^2_j}~,
\label{eq:fdefs}
\end{equation}
and $F^2_0 = \tilde{F}^2_{N+1}=F^2$.

Finally, we will find it useful to denote the (0,0) element of the gauge boson mass matrices as 
\begin{equation}
[M^2_Z]_{0,0} = [M^2_W]_{0,0} = {g^2_0 f^2_1 \over 4} \equiv \tilde{\mathsf m}^2 ~.
\label{eq:gzerofzero}
\end{equation}
To connect with the notation of Ref. \cite{SekharChivukula:2004mu} we note that
\begin{equation}
\tilde{\mathsf m}^{-2} = \Sigma_{[0,1)} \equiv \Sigma_{p=1}
\label{eq:correspond}
\end{equation}


\subsection{Electroweak Parameters}

As we have shown in \cite{Chivukula:2004af}, the most
 general amplitude (to leading order in deviation from the standard model)
 for low-energy four-fermion neutral weak current processes in
any ``universal'' model \cite{Barbieri:2004qk} may be written 
as\footnote{See \cite{Chivukula:2004af} for a discussion of the correspondence 
between the ``on-shell'' parameters defined here, and the zero-momentum
parameters defined in  \protect{\cite{Barbieri:2004qk}}.  Note that $U$ is shown in \cite{Chivukula:2004af} to be zero to the order we consider in this paper.}
\begin{eqnarray}
-{\cal M}_{NC} = e^2 {{\cal Q}{\cal Q}' \over Q^2} 
& + &
\dfrac{(I_3-s^2 {\cal Q}) (I'_3 - s^2 {\cal Q}')}
	{\left({s^2c^2 \over e^2}-{S\over 16\pi}\right)Q^2 +
		{1\over 4 \sqrt{2} G_F}\left(1-\alpha T +{\alpha \delta \over 4 s^2 c^2}\right)
		} 
\label{eq:NC4} \\ \nonumber & \ \ & \\
&+&
\sqrt{2} G_F \,{\alpha \delta\over  s^2 c^2}\, I_3 I'_3 
+ 4 \sqrt{2} G_F  \left( \Delta \rho - \alpha T\right)({\cal Q}-I_3)({\cal Q}'-I_3')~,
\nonumber 
\end{eqnarray}
and the matrix element for charged current process may be written 
\begin{eqnarray}
  - {\cal M}_{\rm CC}
  =  \dfrac{(I_{+} I'_{-} + I_{-} I'_{+})/2}
             {\left(\dfrac{s^2}{e^2}-\dfrac{S}{16\pi}\right)Q^2
             +{1\over 4 \sqrt{2} G_F}\left(1+{\alpha \delta \over 4 s^2 c^2}\right)
            }
        + \sqrt{2} G_F\, {\alpha  \delta\over s^2 c^2} \, {(I_{+} I'_{-} + I_{-} I'_{+}) \over 2}~.
\label{eq:CC3}
\end{eqnarray}
Note that the parameter $s^2$ is defined implicitly in these expressions as the ratio of
the ${\cal Q}$ and $I_3$ couplings of the $Z$ boson.
$S$ and $T$ are the familiar oblique electroweak parameters \cite{Peskin:1992sw,Altarelli:1990zd,Altarelli:1991fk}, 
as determined by examining the {\it on-shell} properties of the $Z$ and $W$ bosons.
$\Delta \rho$ corresponds to the deviation from unity of the ratio of the strengths of
low-energy isotriplet weak neutral-current scattering and charged-current scattering.
Finally, the contact interactions proportional to $\alpha \delta$ and ($\Delta \rho - \alpha T$)
correspond to ``universal non-oblique'' corrections. 

From our previous analysis \cite{SekharChivukula:2004mu}, we know that for
a model of the sort shown in Figure~ \ref{fig:tone}, $\Delta \rho \equiv 0$.
In the limit in which we will work (see eqns. (\ref{eq:smallgs}) and 
(\ref{eq:smallgconsequence})), we will also find (Section 5) that $\alpha T \approx 0$. Therefore our analysis of
electroweak corrections in these models reduces to computing the values
of $\alpha S$ and $\alpha \delta$.

\subsection{The Limit Taken}

We will study 
the correlation functions for $0 \le -Q^2 \le (200\,{\rm GeV})^2$
at tree-level assuming that the heavy $W$- and $Z$-bosons
satisfy
\begin{equation}
{\mathsf m}^2_{Zz},\, {\mathsf m}^2_{Ww} \gg (200\, {\rm GeV})^2~.  \qquad [z,w = 1,...,N]
\label{eq:heavyz}
\end{equation}
From our previous analysis \cite{SekharChivukula:2004mu}, we 
 expect that $g_{N+1}$ (being the only $U(1)$ coupling) must 
be smaller than the other $g_i$ in order to ensure that a light $Z$ state will exist.  
In principle, any one of the $SU(2)$ couplings could also be small 
(to ensure the presence of a light $W$).  However, our previous analysis  \cite{SekharChivukula:2004mu} tells us that, in a viable model, any site 
with small coupling must be linked by large $f$-constants to site 0. For simplicity,
we will therefore restrict our attention to the case where the only $SU(2)$ site with a small
coupling is site 0. This may be viewed
as analyzing the general model after having ``integrated out'' the 
links  with large $f$-constants. 

In our analytic work, therefore, we will
work  in the limit that 
\begin{equation}
g_0,\, g_{N+1} \ll g_i~, \qquad i=1,\ldots,N~.
\label{eq:smallgs}
\end{equation}
From the analyses presented in \cite{SekharChivukula:2004mu}, we find
that in the the limit of eqn. (\ref{eq:smallgs}),
\begin{equation}
\Sigma_Z \approx \Sigma_W \approx \Sigma_{(0,N+1)} \equiv \Sigma_r ~,
\label{eq:smallgconsequence}
\end{equation}
where the last definition makes contact with the notation $M^2_{(0,N+1)} \equiv M^2_r$ in Ref. \cite{SekharChivukula:2004mu}, and
\begin{equation}
M^2_W = {g^2_0 F^2 \over 4} + {\cal O}\left({M^2_W \over {\mathsf m}^2_{W1}}\right)~.
\label{eq:approxmw}
\end{equation}
Note that we expect
$g_0$ to be approximately of order the standard model $SU(2)$ coupling and
therefore numerically of order 1 -- the limit in eqn. (\ref{eq:smallgs})
implies that the other $g_i$ will be larger, and eqn. (\ref{eq:approxmw}) implies
that $F\simeq\, 246$ GeV.

For phenomenologically motivated reasons (see eqn. (\ref{eq:newlimiti})), 
we will also take
\begin{equation}
x_1 {\{-Q^2,\,M^2_W\} \over \tilde{\mathsf m}^2} \ll 1~.
\label{eq:newlimit}
\end{equation}
This approximation may be 
satisfied either by $x_1$ being small,  $\tilde{\mathsf m}^2$ being large, or some
combination thereof.

In the numerical examples studied in Section 8, we calculate the  tree-level 
masses and residues exactly, and we confirm that our analytic calculations based on the approximations
of eqn. (\ref{eq:smallgs}) and (\ref{eq:newlimit}) do indeed capture the essential features of models with one-site delocalization.

\section{\protect{$\mathbf{ [G_{NC}(Q^2)]_{WY}}$}, \protect{$\mathbf{[\xi_Z]_{WY}}$} and \protect{$\mathbf{\alpha S}$}}

We begin by computing $[G_{NC}(Q^2)]_{WY}$.  Starting from eqn. (\ref{eq:gncwyii}), we  see the two
contributions coming from the two sites at which the fermion resides.
Based on Ref. \cite{SekharChivukula:2004mu}, then, we may immediately compute the 
the two relevant elements of the propagator matrix
\begin{eqnarray}
[G_{NC}(Q^2)]_{0,N+1} & = & {e^2 M^2_Z \over Q^2 (Q^2 + M^2_Z)}
\left[ \prod_{z=1}^N {{\mathsf m}^2_{Zz} \over Q^2 + {\mathsf m}^2_{Zz}}\right] \cr
[G_{NC}(Q^2)]_{1,N+1} & = & {e^2 M^2_Z \over Q^2 (Q^2 + M^2_Z)}
\left[ { Q^2+ \tilde{\mathsf m}^2 \over \tilde{\mathsf m}^2}\right]
\left[ \prod_{z=1}^N {{\mathsf m}^2_{Zz} \over Q^2 + {\mathsf m}^2_{Zz}}\right]~.
\label{eq:usefularray}
\end{eqnarray}
Combining these results, we find
\begin{equation}
[G_{NC}(Q^2)]_{WY} = {e^2 M^2_Z \over Q^2 (Q^2 + M^2_Z)}
\left[ 1+ x_1{ Q^2 \over \tilde{\mathsf m}^2}\right]
\left[ \prod_{z=1}^N {{\mathsf m}^2_{Zz} \over Q^2 + {\mathsf m}^2_{Zz}}\right]~.
\label{eq:gncwyi}
\end{equation}
Given eqns. (\ref{eq:heavyz}) and (\ref{eq:newlimit}), we may expand the final product in this
expression and find
\begin{equation}
[G_{NC}(Q^2)]_{WY} = {e^2 M^2_Z \over Q^2 (Q^2 + M^2_Z)}
\left[ 1+ Q^2({ x_1\over \tilde{\mathsf m}^2}-\Sigma_Z)\right]~.
\label{eq:gncwy}
\end{equation}
If we take
\begin{equation}
x_1 = {\Sigma_Z \over \Sigma_{[0,1)}} = \tilde{\mathsf m}^2 \Sigma_Z~.
\label{eq:sbar}
\end{equation}
we have that (in this momentum range) this correlation function equals its standard model
value to leading order,  
\begin{equation}
[G_{NC}(Q^2)]_{WY} \equiv [G_{NC}(Q^2)]^{SM}_{WY}~.
\end{equation}

Next, we compute $[\xi_Z]_{WY}$, from which we may directly
extract $\alpha S$.  The residue decomposes like the correlation function
\begin{equation}
[\xi_Z]_{WY} = x_0 [\xi_Z]_{0,N+1} + x_1[\xi_Z]_{1,N+1}~,
\end{equation}
where the subscripts on the right hand side of the equation denote
the residue of the pole of the corresponding propagator matrix element.
From eqn. (\ref{eq:usefularray}), we 
find
\begin{eqnarray}
[\xi_Z]_{0,N+1} & = & -e^2[1+M^2_Z \Sigma_Z] \cr
[\xi_Z]_{1,N+1} & = & -e^2[1+M^2_Z(\Sigma_Z - \Sigma_{[0,1)})]~.
\end{eqnarray}
Combining these results, we find
\begin{equation}
[\xi_Z]_{WY} = -e^2[1+M^2_Z(\Sigma_Z - x_1 \Sigma_{[0,1)})] ~.
\label{eq:xizwy}
\end{equation}
The form of the four-fermion weak interaction amplitudes of eqns. (\ref{eq:NC4})
and (\ref{eq:CC3}) implies  \cite{SekharChivukula:2004mu}
\begin{equation}
   \dfrac{1}{e^2}[\xi_Z]_{WY} =     -1 -  \dfrac{\alpha}{4s_Z^2 c_Z^2} S~,
\end{equation}
and hence we find
\begin{equation}
\alpha S = 4 s^2_Z c^2_Z M^2_Z (\Sigma_Z - x_1 \tilde{\mathsf m}^{-2})~.
\label{eq:alphas}
\end{equation}
Now it is clear that the same ``tuning'' of the localization of the fermion in conjunction with the heavy
$Z$-boson mass matrix that causes $[G_{NC}(Q^2)]_{WY}$ to have its standard model form at low momentum likewise makes $\alpha S$ small. In fact, if equation (\ref{eq:sbar}) is satisfied, 
then $\alpha S \simeq 0$ 

Using eqn. (\ref{eq:correspond}), we may rewrite this result in the form
\begin{equation}
\alpha S = 4 s^2_Z c^2_Z M^2_Z (\Sigma_Z - x_1 \Sigma_{p=1})~,
\label{eq:alphasi}
\end{equation}
which agrees with the results of \cite{SekharChivukula:2004mu} when
$x_0=1$ or $x_1=1$, and smoothly interpolates between these extremes.

Finally, note that, in order for $\alpha S$ to be small, we need
\begin{equation}
x_1\,{M^2_W \over \tilde{\mathsf m}^2} = M^2_W \Sigma_Z \ll 1~,
\label{eq:newlimiti}
\end{equation}
and the limit of eqn. (\ref{eq:newlimit}) is directly related to that
of eqn. (\ref{eq:heavyz}).

\section{\protect{$\mathbf{[G_{NC}(Q^2)]_{YY}}$}, \protect{$\mathbf{[\xi_Z]_{YY}}$} and \protect{$\mathbf{\alpha T}$}}

Next, consider the correlation function $[G_{NC}(Q^2)]_{YY}$. Given the structure
of the moose in Figure~ \ref{fig:tone} and the form of the fermion couplings
in eqn. (\ref{eq:fcoupling}), we see that the delocalization of the fermions is
irrelevant in this case -- we get the same result \cite{SekharChivukula:2004mu} as in the case of the simplest Case I
model:
\begin{equation}
[G_{NC}(Q^2)]_{YY} = {e^2 M^2_Z (Q^2+M^2_W) \over Q^2 M^2_W (Q^2 + M^2_Z)}
\left[\prod_{w=1}^N {Q^2 + {\mathsf m}^2_{Ww} \over {\mathsf m}^2_{Ww}}\right]
\left[ \prod_{z=1}^N {{\mathsf m}^2_{Zz} \over Q^2 + {\mathsf m}^2_{Zz}}\right]
~.
\label{eq:gncyyi}
\end{equation}
Expanding for low $Q^2$ (see eqn. (\ref{eq:heavyz})) we find, to this order,
\begin{equation}
[G_{NC}(Q^2)]_{YY} = {e^2 M^2_Z (Q^2+M^2_W) \over Q^2 M^2_W (Q^2 + M^2_Z)}
\left[1+Q^2(\Sigma_W - \Sigma_Z)\right] = [G_{NC}(Q^2)]^{SM}_{YY}~.
\end{equation}
where the last equality follows from eqn. (\ref{eq:smallgconsequence}) and where
$[G_{NC}(Q^2)]^{SM}_{YY}$ denotes the tree-level standard model value in terms of $e^2$,
$M^2_W$, and $M^2_Z$ .

The residue is likewise the same as in the simplest Case I model: 
\begin{equation}
[\xi_Z]_{YY} = {e^2 ( M^2_Z - M^2_W) \over M^2_W}\,
\left[1+M^2_Z(\Sigma_Z - \Sigma_W) \right]~.
\label{eq:xizyy}
\end{equation}
Therefore, using the results of  \cite{SekharChivukula:2004mu},
we find
\begin{equation}
\alpha T = s^2_Z M^2_Z (\Sigma_Z - \Sigma_W) \simeq 0~,
\label{eq:smallT}
\end{equation}
independent of the value of $x_0$. The last equality follows from 
eqn. (\ref{eq:smallgconsequence}) (i.e. from working in the limit $g^2_{N+1} \ll 1$). 

\section{\protect{$\mathbf{[G_{CC}(Q^2)]_{WW}}$}, \protect{$\mathbf{[\xi_W]_{WW}}$}, and 
\protect{$\mathbf{\alpha \delta}$}}

Finally, to compute $\alpha \delta$ we must compute a $WW$ correlation function. For simplicity,
we will consider the charged-current correlation function $[G_{CC}(Q^2)]_{WW}$. 
We may do so by recalling that the matrix $G_{CC}(Q^2)$ is defined by
\begin{equation}
[G_{CC}(Q^2)]_{i,j} \equiv g_i g_j \left[(Q^2+M^2_W)^{-1}\right]_{i,j}~.
\label{eq:sixone}
\end{equation}
The correlation function of $J^+_\mu$ with $J^-_\nu$
is therefore proportional to
\begin{equation}
x_0^2 [G_{CC}(Q^2)]_{0,0} + 2 x_0 x_1  [G_{CC}(Q^2)]_{0,1}
+ x^2_1[G_{CC}(Q^2)]_{1,1}~.
\label{eq:gccww}
\end{equation}

To make progress, we relate the various propagator elements to one another.
Consider eqn.~(\ref{eq:sixone}) as a matrix equation
\begin{equation}
G_{CC}(Q^2) = G \cdot {{\cal I}\over Q^2 + M^2_W} \cdot G~,
\end{equation}
where $G$ is the matrix of gauge coupling constants
and  ${\cal I}$ denotes the identity matrix in gauge space.
From this, we immediately
see that we have the matrix relation
\begin{equation}
(Q^2+M^2_W) \cdot G^{-1} \cdot G_{CC}(Q^2) \cdot G^{-1} \equiv {\cal I}~.
\end{equation}
Applying this
relation explicitly to the first row of the matrix $(Q^2+M^2_W)$ and the first two columns
of the matrix $G_{CC}(Q^2)$, we find the relations\footnote{The propagator matrix elements
$[G_{CC}(Q^2)]_{0,1}$ and $[G_{CC}(Q^2)]_{0,0}$ do not have poles
at $Q^2=-\tilde{\mathsf m}^2$, as might be inferred from the form of eqns.
(\protect\ref{eq:newrelationi}) and (\protect\ref{eq:newrelationii}). Rather, these
potential poles are cancelled by zeros of the numerators in these expressions.}
\begin{equation}
[G_{CC}(Q^2)]_{0,1} =  {\tilde{\mathsf m}^2 \over Q^2 + \tilde{\mathsf m}^2} [G_{CC}(Q^2)]_{1,1}~,
\label{eq:newrelationi}
\end{equation}
and
\begin{equation}
[G_{CC}(Q^2)]_{0,0} =  {g^2_0 \left(1+{f^2_1 \over 4}[G_{CC}(Q^2)]_{0,1}\right)\over 
Q^2 + \tilde{\mathsf m}^2}~.
\label{eq:newrelationii}
\end{equation}

Using these results, we find
\begin{equation}
[G_{CC}(Q^2)]_{WW} = \left(1+x_1\,{Q^2\over \tilde{\mathsf m}^2}\right)^2[G_{CC}(Q^2)]_{0,0}
-{8x_1\over f^2_1} + {4 x^2_1 \over f^2_1}\left(1-{Q^2\over \tilde{\mathsf m}^2}\right)~.
\label{eq:sixseven}
\end{equation}
Given the limit of eqn. (\ref{eq:newlimit}), for the momenta of interest
this  reduces to 
\begin{equation}
[G_{CC}(Q^2)]_{WW} = \left(1+2 x_1\,{Q^2\over \tilde{\mathsf m}^2}\right)[G_{CC}(Q^2)]_{0,0}
-{8x_1\over f^2_1} + {4 x^2_1 \over f^2_1}~.
\label{eq:gccqsq}
\end{equation}
We can re-arrange this to isolate the pole at $Q^2=-M^2_W$ from the non-pole pieces of the correlation function:
\begin{eqnarray}
[G_{CC}(Q^2)]_{WW} & = & \left(1-2 x_1\,{M^2_W\over \tilde{\mathsf m}^2}\right)[G_{CC}(Q^2)]_{0,0} + {2x_1 \over \tilde{\mathsf m}^2}
(Q^2+M^2_W) [G_{CC}(Q^2)]_{0,0} \cr
& - & {4 x_1\over F^2}  \left({M^2_W \over \tilde{\mathsf m}^2}\right)
(2-x_1)~,
\label{eq:isolate}
\end{eqnarray}
where we have used eqn. (\ref{eq:approxmw}) to simplify the last term.

From the pole term (first term) of eqn. (\ref{eq:isolate}) we see that the  residue of the charged-current correlation function at 
$Q^2=-M^2_W$ is
\begin{equation}
[\xi_W]_{WW} =  \left(1-2 x_1\,{M^2_W\over \tilde{\mathsf m}^2}\right) [\xi_W]_{0,0}~.
\label{eq:xiwwwi}
\end{equation}
Applying the calculations presented in \cite{SekharChivukula:2004mu}, 
we observe that
\begin{equation}
[\xi_W]_{0,0} = {e^2 M^2_Z \over M^2_Z - M^2_W}
\left[1+M^2_W (\Sigma_Z + \Sigma_W)\right] =
[\xi_W]^{SM} \left[1+2 M^2_W \Sigma_Z \right]~,
\end{equation}
where the last equality follows from eqn. (\ref{eq:smallgconsequence}),
and $[\xi_W]^{SM}$ denotes the tree-level standard model value of the residue
expressed in terms of $M^2_{W,Z}$. Therefore, we find from  eqn. (\ref{eq:xiwwwi}) that
\begin{equation}
[\xi_W]_{WW} =  \left(1-2 x_1\,{M^2_W\over \tilde{\mathsf m}^2} +2 M^2_W \Sigma_Z \right)
[\xi_W]^{SM} 
\end{equation}
For $x_1=\tilde{\mathsf m}^2 \Sigma_Z$ (i.e., for $\alpha S = 0$), the residue of the pole equals the standard model value.  This is consistent with the form of  eqn. (\ref{eq:CC3}).

While the residue of the $W$-pole is given by its standard model value,
the non-pole terms in eqn. (\ref{eq:isolate}) give rise to a non-zero
value of $\alpha \delta$.
From the analyses presented in \cite{SekharChivukula:2004mu}, 
\begin{equation}
[G_{CC}(Q^2)]_{0,0} =  {4 M^2_W  \over  F^2 [Q^2 + M^2_W]}
\left[\prod_{w=1}^N {{\mathsf m}^2_{Ww} \over Q^2 +{\mathsf m}^2_{Ww}}\right]
\left[\prod_{r=1}^N {Q^2 + {\mathsf m}^2_r \over {\mathsf m}^2_r}\right]~.
\end{equation}
Expanding the product for the momenta of interest, this may be written
\begin{equation}
[G_{CC}(Q^2)]_{0,0} = {4  M^2_W \over  F^2 [Q^2 + M^2_W]}
\left[1+Q^2 (\Sigma_r - \Sigma_W)\right]
\approx {4  M^2_W \over  F^2 [Q^2 + M^2_W]}~,
\end{equation}
where the last equality follows from eqn. (\ref{eq:smallgconsequence}).
Comparing the non-pole terms in eqn. (\ref{eq:isolate}) with the
form of the matrix element eqn. (\ref{eq:CC3}), we therefore calculate
\begin{equation}
\sqrt{2} G_F {\alpha \delta \over s^2 c^2} = {4  x^2_1 \over F^2} 
\left({M^2_W \over \tilde{\mathsf m}^2}\right)~.
\label{eq:gfff}
\end{equation}
However, 
\begin{equation}
\sqrt{2} G_F  \equiv  {1\over 4} [G_{CC}(Q^2=0)]_{WW} 
\end{equation}
so from eqn. (\ref{eq:gccqsq}), again using eqns. (\ref{eq:gzerofzero})
and (\ref{eq:approxmw}), we see that
\begin{equation}
\sqrt{2} G_F =  {1\over F^2} -{(2-x_1)x_1\over f^2_1} \approx {1\over F^2}\left[1-{(2-x_1)x_1 M^2_W \over \tilde{\mathsf m}^2}\right] 
=   {1\over F^2}\left[1+{\cal O}\left({x_1 M^2_W \over \tilde{\mathsf m}^2}\right)\right]
\end{equation}
Using this in eqn. (\ref{eq:gfff}) we find
\begin{equation}
{\alpha \delta \over 4 s^2 c^2} =  x^2_1 {M^2_W \over \tilde{\mathsf m}^2}
\end{equation}
If we employ eqn. (\ref{eq:correspond}), this can be rewritten as 
\begin{equation}
{\alpha \delta \over 4 s^2 c^2} =  x^2_1 
M^2_W \Sigma_{p=1}~,
\label{eq:adeltaeqn}
\end{equation}
which agrees with the results of  \cite{SekharChivukula:2004mu} for
$x_0=1$ or $x_1=1$ and smoothly interpolates between them.

When we choose the amount of delocalization to ensure that $\alpha S$ vanishes, $x_1 = \tilde{\mathsf m}^2 \Sigma_Z$,  we find
\begin{equation}
{\alpha \delta \over 4 s^2 c^2} = x_1\, M^2_W \Sigma_Z~.
\label{eq:alphadelta}
\end{equation}
Moreover,  as argued previously \cite{SekharChivukula:2004mu}, preserving the unitarity of 
longitudinal $W$ boson scattering requires that $M^2_W \Sigma_Z
\ge 4 \times 10^{-3}$. The results of \cite{Barbieri:2004qk} imply\footnote{When $\alpha
S=0=(\Delta \rho - \alpha T) $, from eqn. (\protect\ref{eq:final}) we see that 
$\hat{S} = \alpha \delta /4 s^2 c^2$.}
that experiment currently imposes the upper bound $\alpha \delta/4s^2 c^2 < 1 \times 10^{-3}$ at 95\% CL.  Hence we must have
\begin{equation}
x_1 \le {1\over 4}~,
\end{equation}
and the amount of delocalization is bounded to be less than of order 25\%.

\section{Beyond Extra Dimensions}

\label{sec:beyond}

\subsection{Re-interpreting fermion delocalization}

The ``delocalized'' fermion coupling in deconstructed Higgsless models, 
eqn. (\ref{eq:fcoupling}), may also be written using the Goldstone
boson fields of the Moose in Figure~ \ref{fig:tone}. Consider the current operator
\begin{equation}
{\rm Tr} \left( {\sigma^a\over 2} U^\dagger_1 iD_\mu U_1 \right) \rightarrow
+{1\over 2}(A^a_{0\mu} - A^a_{1\mu})~,
\end{equation}
where the $\sigma$ are the Pauli matrices, $D_\mu$ is the covariant derivative
\begin{equation}
i\, D_\mu U_1 = i\, \partial_\mu U_1 + {\vec{A}_{0\mu} \cdot \vec{\sigma}\over 2}U_1
-  U_1 {\vec{A}_{1\mu} \cdot \vec{\sigma}\over 2}~,
\end{equation}
consistent with eqn. (\protect\ref{eq:fcoupling}), 
and where  we have specified the form of this operator in unitary gauge, where all the link
fields $U_j \equiv {\cal I}$.
In this language, we see that the fermions' weak couplings in eqn. (\ref{eq:fcoupling}) 
may be written
\begin{equation}
\vec{J}^\mu_L \cdot \left[ \vec{A}^0_\mu -  2 x_1
{\rm Tr} \left( {\vec{\sigma}\over 2} U^\dagger_1 iD_\mu U_1 \right) \right]~.
\label{eq:newcoupling}
\end{equation}
From this point of view, the fermions are charged only under $SU(2)_0$ and
the apparent delocalization comes about from couplings to the Goldstone-boson fields.\footnote{This also generalizes naturally to models with multi-site delocalization.}

Note that, in the gauge-boson normalization we are using, the linear combination of
gauge fields $A^a_{0\mu} - A^a_{1\mu}$ are strictly orthogonal to the photon
\begin{equation}
A^\gamma_{\mu} \propto A^3_{0\mu} + A^3_{1\mu} +\ldots + A^3_{N+1\, \mu}~.
\end{equation}
Hence, the couplings of eqn. (\ref{eq:newcoupling}) result in a modification of the
$Z$ and $W$-couplings whose size depends on the $x_1$ and the admixture of $A_0-A_1$
in the mass-eigenstate $W$ and $Z$ fields. {\it But the couplings of eqn. (\ref{eq:newcoupling}) do not modify the photon coupling.}

\EPSFIGURE[v]{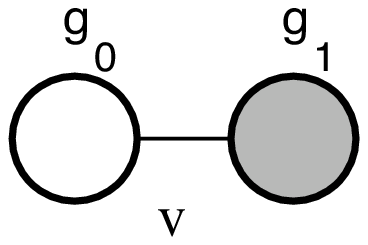,width=0.2\textwidth}
{Simple two-site moose diagram corresponding 
to the global symmetry structure of the one-Higgs doublet 
standard model or the simplest one-doublet technicolor model.
\label{fig:toneed}}

\subsection{Technicolor}

We have seen that fermion delocalization, on the one hand, affects $\alpha S$,
and, on the other, can be rewritten as the fermions coupling to the Goldstone boson
currents. We can apply the idea of fermions' coupling to Goldstone bosons
directly to technicolor -- a two-site model -- which has no extra-dimensional
interpretation.
Consider the two-site model of Figure~ \ref{fig:toneed}, with fermion couplings\footnote{This
kind of operator was previously considered in references 
\protect\cite{Hall:1986wx,Chivukula:1987zq,Randall:1989ds,Chivukula:1992ap},
the first of these prior to the definition of $\alpha S$ and the last considering
only flavor-dependent effects. Note that there is only one $SU(2)$ group.}
\begin{equation}
{\cal L}_f = \vec{J}^\mu_L \cdot \left[ \vec{A}^0_\mu - 
2x_1\,{\rm Tr} \left( \Sigma^\dagger {\vec{\sigma}\over 2} \, iD_\mu \Sigma\right)\right]
+ J^\mu_Y A^1_\mu~,
\label{eq:tcoperator}
\end{equation}
where $\Sigma$ is the unitary matrix representing the three eaten Goldstone bosons,
and the $\vec{\sigma}$ are the Pauli matrices.
Following \cite{Chivukula:1992ap}, we find that in unitary gauge
\begin{equation}
2x_1 \vec{J}^\mu_L \cdot {\rm Tr} \left( \Sigma^\dagger{\vec{\sigma}\over 2} \, iD_\mu \Sigma \right)
\rightarrow 
-2x_1\,\left[ {e \over \tilde{s} \tilde{c}} Z^\mu J^3_\mu + {e\over \tilde{s} \sqrt{2}}
\left(W^{+\mu}J^-_\mu + W^{-\mu} J^+_\mu \right) \right]~,
\label{eq:unitary}
\end{equation}
where
\begin{equation}
g_0 = {e \over \tilde{s}}~, \qquad g_1 = {e\over \tilde{c}}~.
\end{equation}
Hence, we find the overall $Z$ and $W$ couplings
\begin{equation}
{e\over \tilde{s}\tilde{c}}\,Z^\mu
\left[(1-2x_1)J^3_\mu-\tilde{s}^2 J^{\cal Q}_\mu \right]~,
\label{eq:shiftzppp}
\end{equation}
\begin{equation}
{e\over \sqrt{2} \tilde{s}} \left(1-2 x_1\right)
\left[W^{+\mu} J^-_\mu + W^{-\mu} J^+_\mu \right]~.
\label{eq:shiftwpp}
\end{equation}
Comparing with eqns. (\ref{eq:NC4}) and (\ref{eq:CC3}) we find
\begin{equation}
\tilde{s}^2 = s^2(1- 2x_1)~, \qquad \alpha\,\Delta S \approx -\, 8 s^2 x_1~,
\label{eq:stilde}
\end{equation}
As anticipated, the Goldstone-boson operator in eqn. (\ref{eq:tcoperator}) can shift $\alpha S$.  In fact, it will
shift $\alpha S$ in a negative direction (since $x_1$ is positive) just as occurs in eqn. (\ref{eq:alphas}).

In a technicolor model this effect could be used to cancel the positive QCD-size
value of $\alpha S$ arising \cite{Peskin:1992sw,Golden:1990ig,Holdom:1990tc}
from the $L_{10}$ operator. It is also amusing to note that the sign of $x_1$ arising from
the ETC operators considered in \cite{Chivukula:1992ap} is positive. If the operator
of eqn. (\ref{eq:tcoperator}) arises from ETC exchange, that reference found (note that
the convention for the covariant derivative differs in that reference)
\begin{equation}
x_1 = {\xi^2\over 4} {g^2_{ETC} v^2 \over M^2_{ETC}}~,
\end{equation}
where $g_{ETC}$ and $M_{ETC}$ are the extended technicolor coupling and gauge-boson
mass, $v \simeq 246$ GeV, and $\xi$ is a model-dependent Clebsch-Gordon coefficient.
The canonical QCD-like technicolor estimate gives $S_{TC} = {\cal O}(0.5)$. If we require the
sum of the canonical contribution plus that arising from eqn. (\ref{eq:stilde}) to vanish, we find
$|x_1| \simeq 2\times 10^{-3}$, and hence
\begin{equation}
{M_{ETC} \over \xi\, g_{ETC}} \simeq 3\,{\rm TeV}~.
\end{equation}
In an ETC model, one would also expect contributions to $\alpha \delta$ (from
ETC exchange) of the same order  -- which 
implies there must be a Clebsch of order a few to suppress $\alpha \delta$
relative to $x_1$ \cite{Barbieri:2004qk}. One could imagine, for example, a model
with flavor-independent left-handed low-scale ETC interactions (with light quark masses
suppressed by high-scale right-handed interactions).

\section{Examples of Delocalized Deconstructed Higgsless Models}

\EPSFIGURE[h]{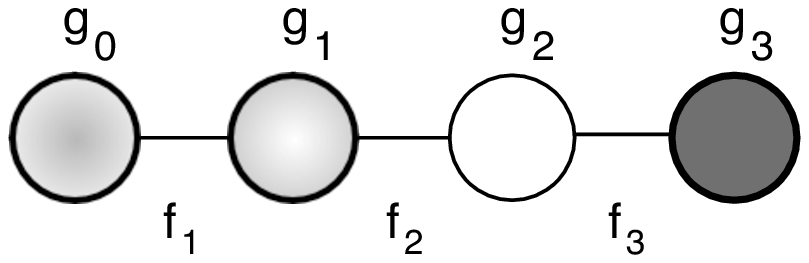,width=8cm}
{The model analyzed in the explicit numerical calculation.
  Sites 0 to 2 are $SU(2)$ gauge groups, while site 3 is $U(1)$.
  Fermions are coupled to sites 0 and 1 (weak isospin), and to site 3
  (weak hypercharge), as denoted by the thick circles. \label{fig:4sites}}

To illustrate the ideas discussed in the earlier sections of the paper, we now study
 a linear moose model with 4 sites and 3 links (Figure ~\ref{fig:4sites}), a model small enough to be easily solved numerically without approximations. 
We will calculate the  tree-level 
masses and residues exactly and confirm that our previous analytic calculations based on the approximations
of eqn. (\ref{eq:smallgs}) and (\ref{eq:newlimit}) capture the essential features of models with one-site delocalization.

Starting from this $[SU(2)]^3\times U(1)$ gauge structure, we introduce a chiral fermion $\psi_{0L}$ (assumed to be a
doublet of $SU(2)_0$), and a Dirac fermion $\psi_1 = \psi_{1L} +
\psi_{1R}$ (doublet of $SU(2)_1$).  Both 
$\psi_{0L}$ and $\psi_{1}$ are assumed to have the same weak
hyperchage $Y_\psi$.
The fermion sector of this model is then given by the Lagrangian,
\begin{eqnarray}
  {\cal L}_{\rm fermion}
  &=&
    \bar{\psi}_{0L} 
    (i\fsl{\partial} + \frac{\tau^a}{2} \fsl{A}^a_0 + Y_\psi \fsl{A}_3
    ) \psi_{0L}
  \nonumber\\
  & & 
   +\bar{\psi}_{1L} 
    (i\fsl{\partial} + \frac{\tau^a}{2} \fsl{A}^a_1 + Y_\psi \fsl{A}_3
    ) \psi_{1L}
   +\bar{\psi}_{1R} 
    (i\fsl{\partial} + \frac{\tau^a}{2} \fsl{A}^a_1 + Y_\psi \fsl{A}_3
    ) \psi_{1R}.
\end{eqnarray}
The fermion mass term consistent with the gauge symmetry is given by
\begin{equation}
  {\cal L}_{\rm mass} = 
  (\bar{\psi}_{0L}, \bar{\psi}_{1L})
  \left(
    \begin{array}{c}
      y_\psi f_1 U_1 \\
      M_{\psi}
    \end{array}
  \right) \psi_{1R} 
  + \mbox{h.c..}
\end{equation}
After the gauge symmetry breaking, $\psi_{1R}$ and
\begin{equation}
  \psi_{L}^{(1)} = s_\psi \psi_{0L} + c_\psi \psi_{1L} 
\end{equation}
form a Dirac fermion and become massive; we identify this as a KK mode.  There also 
 remains a massless fermion 
\begin{equation}
  \psi_{L}^{(0)} = c_\psi \psi_{0L} - s_\psi \psi_{1L}\ ,
\end{equation}
where 
\begin{equation}
  c_\psi \equiv \dfrac{M_\psi}{\sqrt{y_\psi^2 f_1^2 + M_\psi^2}}, \qquad
  s_\psi \equiv \dfrac{y_\psi f_1}{\sqrt{y_\psi^2 f_1^2 + M_\psi^2}}, 
\end{equation}
which we identify  as the standard model fermion.  Then 
$\psi_{L}^{(0)}$ couples to the gauge fields as
\begin{equation}
  {\cal L}_{\rm fermion}
  = 
    \bar{\psi}^{(0)}_{L} 
    (i\fsl{\partial} + x_0  \frac{\tau^a}{2} \fsl{A}^a_0 
                     +x_1  \frac{\tau^a}{2} \fsl{A}^a_1 
                     + Y_\psi \fsl{A}_3
    ) \psi_{L}^{(0)} + \cdots~,
\end{equation}
where $x_0 = c_\psi^2$ and $x_1=s_\psi^2 $.

In our phenomenological calculations, we use $\alpha, G_F$, and $M_Z$  to specify the input parameters of the standard model. \footnote{Note that the tree level value of $M_W$ in this scheme ($ \left. M_W  \right|_{\rm tree} 
\equiv c_Z M_Z = 79.9607 \mbox{ GeV}$) differs from the observed value ($
  \left. M_W  \right|_{\rm exp} = 80.425 \pm 0.038 \mbox{ GeV}$), 
indicating the importance of one-loop radiative correction at 1\% level.
In this paper, however, we restrict ourselves to tree-level.
We thus denote $M_W^{\rm SM} \equiv \left. M_W \right|_{\rm tree} $
in our calculations, and compare all correlation functions to the corresponding tree-level
standard model results.}  The specific values used are \cite{Eidelman:2004wy}
$\alpha^{-1}  = 128.91 \pm 0.02$, $M_Z = 91.1876 \pm 0.0021\mbox{ GeV}$, and 
$G_F = 1.16637 \times 10^{-5} \mbox{ GeV$^{-2}$}$. 
The Weinberg angle in this scheme is defined by
\begin{equation}
  s_Z^2 c_Z^2 \equiv \dfrac{e^2}{4\sqrt{2} G_F M_Z^2}, \qquad
  c_Z^2 \equiv 1 - s_Z^2~,
\label{eq:s_Z_def}
\end{equation}
yielding $ s_Z^2 = 0.23108 \pm 0.00005$.

Our 4-site linear moose model with one delocalized fermion can be
specified by 8 parameters: $f_i$ ($i=1,2,3$), $g_i$ ($i=0,1,2,3$)
and $x_1^2$.  Three combinations of these parameters have values set by the inputs
 $\alpha, G_F$, and $M_Z$.  For instance,
\begin{equation}
  \dfrac{1}{4\pi\alpha} = \sum_{i=0}^3 \dfrac{1}{g_i^2}\ ,
\end{equation}
and 
\begin{equation}
  4\sqrt{2} G_F 
  = [G_{\rm CC}(Q^2=0)]_{WW}~,
  \end{equation}
where one applys eqn. (\ref{eq:usefulsum}) together with 
\begin{equation}
  [G_{\rm CC}(Q^2=0)]_{0,0} =
  \sum_{i=1,2,3} \frac{4}{f_i^2},  \qquad 
  [G_{\rm CC}(Q^2=0)]_{0,1} = 
  [G_{\rm CC}(Q^2=0)]_{1,1}  =
  \sum_{i=2,3} \frac{4}{f_i^2},
\end{equation}
Requiring $S=0$ sets the value of one more combination, as in eqn. (\ref{eq:newlimiti}).
In order to specify the remaining  parameters, we  adopt three ansatzes 
\begin{equation}
  f_2 = f_3, \quad g_1 = g_2 = 4~.
\end{equation}
The ansatz $f_2 = f_3$ allows us to maximize the delay of the onset of unitarity
violation in longitudinal $W$ scattering.
The large values of $g_1$ and $g_2$ are taken so as to push up the
mass of the gauge-boson  KK-modes. 
Combining the four requirements from $(\alpha, G_F, M_Z, S)$ with the three
ansatzes, only one free parameter, which we identify as $f_1$, is left
in the 4-site model.

We have analyzed the 4-site model with three sample values of the single free parameter: $f_1=300$ GeV (Set 1), $f_1=1000$ (Set 2)
and $f_1= 2000$ GeV (Set 3).  Once $f_1$ is chosen, the other $f_i$, the $g_i$ and $x_1$ have values given in the Table below, as set by the four inputs and three ansatzes.  The masses listed as outputs in the Table were calculated by diagonalizing the 
gauge-boson mass-squared matrix numerically.
\begin{center}
\begin{tabular}{| l || l | l | l | }
\hline 
  & Set 1 & Set 2& Set 3\\
\hline 
Inputs & & & \\
\hline
$f_1$ & 300 GeV & 1000 GeV &2000 GeV\\
$f_2 = f_3$ & 591.850 GeV& 356.303 GeV & 348.922 GeV  \\
$g_0$ & 0.657164& 0.664421 & 0.663478 \\
$g_1 = g_2$ & 4.0 & 4.0 & 4.0\\
$g_3$ & 0.357650& 0.356505 & 0.356651  \\
$x_1$ & 0.014771& 0.139231 & 0.480892 \\
\hline
Calculated Physical Masses & &  & \\
\hline
$M_W$ & 79.9599 GeV& 79.9486 GeV & 79.9080 GeV  \\
$m_{Z1}$ & 892.459 GeV&976.990 GeV & 983.725 GeV \\
$m_{W1}$ & 888.827 GeV& 975.913 GeV & 982.737 GeV \\
$m_{Z2}$ & 1944.08 GeV& 2162.17 GeV & 4114.49 GeV \\
$m_{W2}$ &  1943.39 GeV& 2162.17 GeV & 4144.49 GeV \\
\hline
\end{tabular}
\end{center}

The calculated value of $M_W$ in Sets 1 and 2 
agrees with the tree level standard model value within the uncertainty of 
$\left. M_W \right|_{\rm exp}$ about $0.038$ GeV.  Hence the measured value
of $M_W$ does not currently exclude either Set 1 or 2. 
 The calculated value of $M_W$ in Set 3
deviates from the tree level standard model value
by about 1.4$\sigma$; Set 3  is therefore marginally excluded.

\subsection{Correlation functions}

To explore the expectations that the electroweak correlation functions will 
resemble their standard model counterparts at low momentum, we calculated their values over the 
 LEP energy range $\sqrt{-Q^2}=90$--$200$ GeV,  and compared the tree-level results 
 in the moose model to the tree-level
 results in the standard model.   We computed $[G_{NC}(Q^2)]_{WY}$ for all three sets of parameters using 
 expression (\ref{eq:gncwyi}) and found no discernible deviation
 of the ratio  $[G_{NC}(Q^2)]_{WY}^{Higgsless}$ / $[G_{NC}(Q^2)]^{SM}_{WY}$ from one.  
 This is consistent with the fact that the model parameters were chosen to make $\alpha S$ vanish.  
 Small deviations from one were found in the corresponding ratios for $[G_{NC}(Q^2)]_{YY,WW}$.

Figure~\ref{fig:YY} depicts the behavior of 
$[G_{\rm NC}(Q^2)]_{YY}/[G_{\rm NC}(Q^2)]_{YY}^{\rm SM}$ as calculated using eqn. (\ref{eq:gncyyi}).  
Set 1, shown by the lowest curve in the figure, 
 is indistinguishable from the standard model.The middle curve shows the ratio for the moose  
with Set 2 parameters; the upper curve shows the effect of using Set 3 parameters instead.   For Sets 2 and 3, 
 the deviation from the standard model value is quite small; the visible deviation near 
$\sqrt{-Q^2}\simeq 90$ GeV comes from the difference between $c_Z M_Z$
and $M_W$. 

\DOUBLEFIGURE[htb]{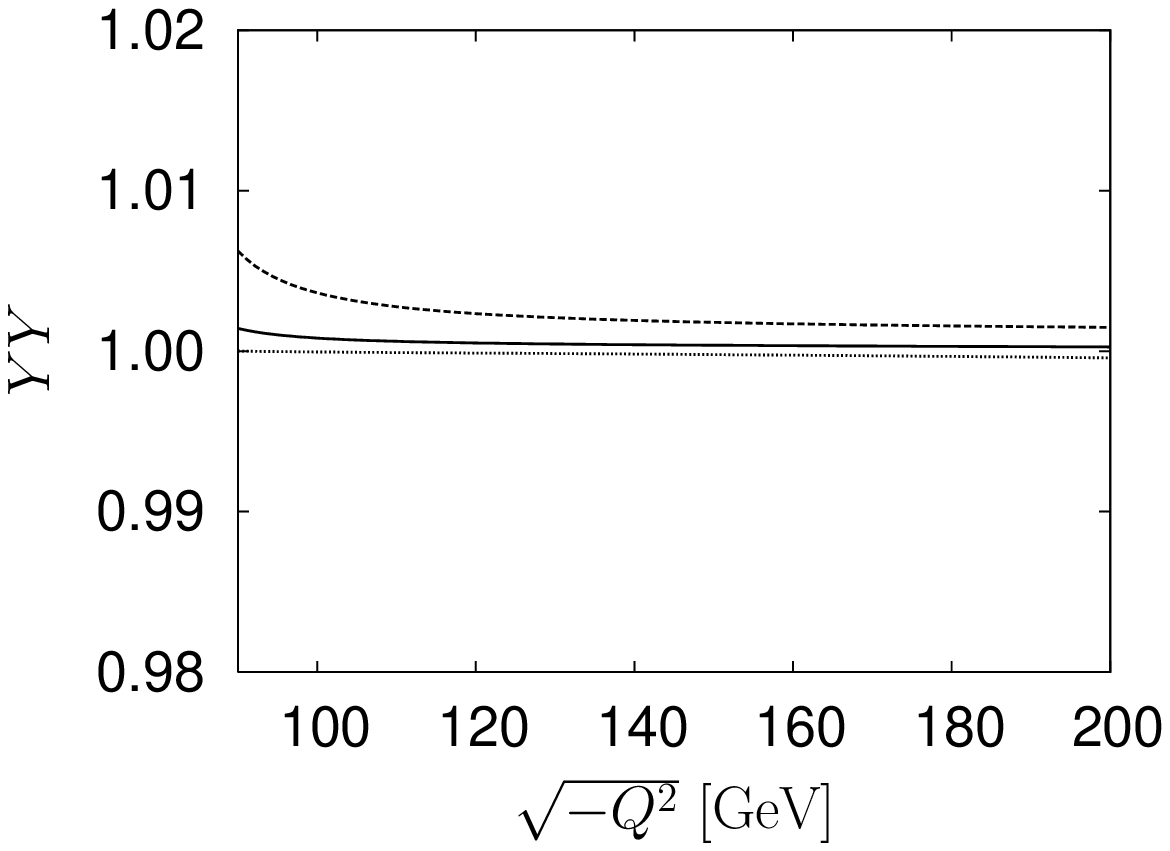,width=8cm}{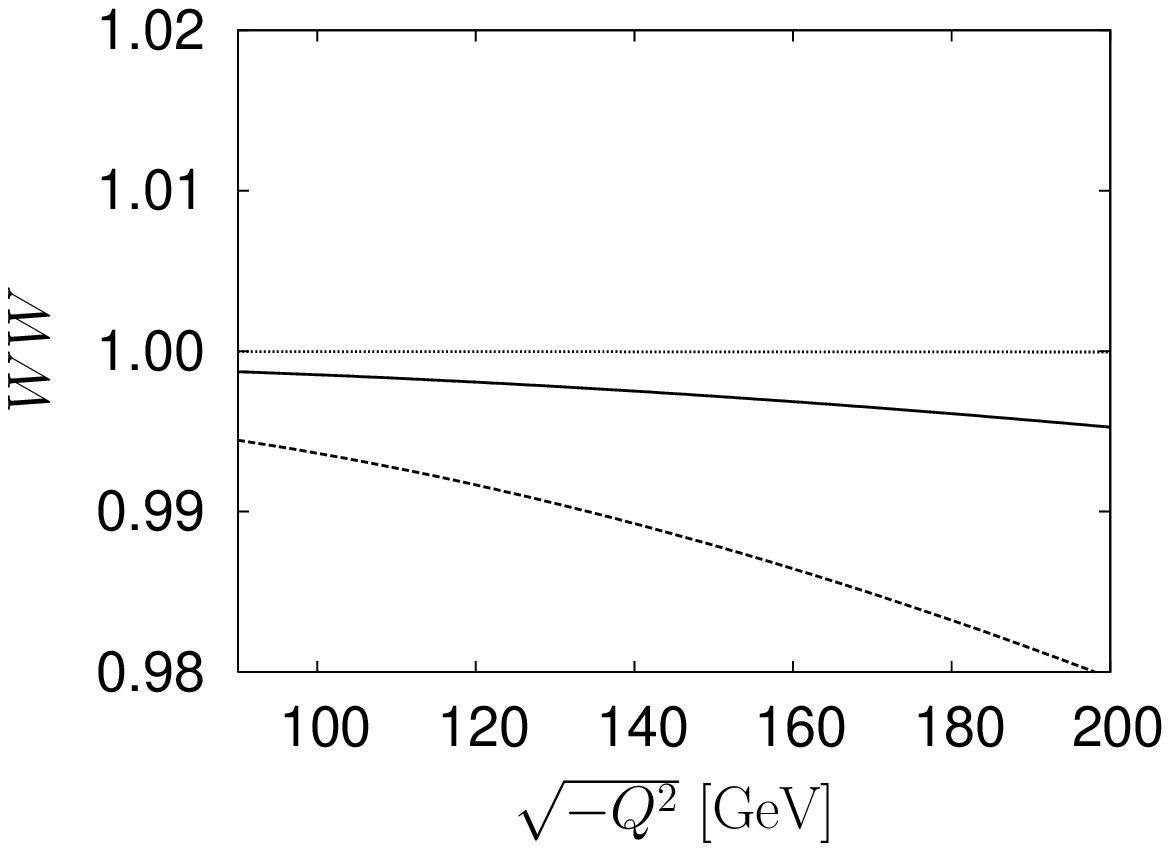,width=8cm}
{$[G_{\rm NC}(Q^2)]_{YY}/[G_{\rm NC}(Q^2)]_{YY}^{\rm SM}$
  for the LEP energy range.  From lowest to highest, the curves are for the Set 1, 2, and
  3. parameters. Set 1 is indistinguishable from the standard model in this plot. \label{fig:YY}}
 {$[G_{\rm NC}(Q^2)]_{WW}/[G_{\rm NC}(Q^2)]_{WW}^{\rm SM}$
  for the LEP energy range.  From highest to lowest, the curves are for the Set 1, 2, and 3 parameters. Set 1 is indistinguishable from the standard model in this plot. \label{fig:WWtex}}

The form of the correlation function $[G_{NC}(Q^2)]_{WW}$ is derived by starting from eqn (\ref{eq:usefulsum}) and working in parallel with the arguments in section 6 and Ref. \cite{SekharChivukula:2004mu} to find 
\begin{eqnarray}
  {}[G_{\rm NC}(Q^2)]_{0,i}
  &=& \dfrac{e^2 M_Z^2}{Q^2(Q^2+M_Z^2)}
      \left[ 
        \prod_{z=1,2} \dfrac{\mathsf{m}_z^2}{Q^2+\mathsf{m}_z^2}
      \right]
      \dfrac{\det \left[ Q^2 + M^2_{(i,3]}\right]}
            {\det \left[M^2_{(i,3]} \right]},
  \\
   {}[G_{\rm NC}(Q^2)]_{1,1}
  &=& 
      \left[
        1 + \dfrac{Q^2}{\tilde\mathsf{m}^2}
      \right] [G_{\rm NC}(Q^2)]_{0,1}
\end{eqnarray}
We calculate the eigenvalues of matrix $M^2_{(0,3]}$ in the four-site model to be
\begin{eqnarray}
 & {\rm Set\  1:} & \quad \quad 43.4066\mbox{ GeV}, \quad
  890.794\mbox{ GeV}, \quad
  1943.92\mbox{ GeV} \\
 & {\rm Set\ 2:} & \quad \quad  43.4996\mbox{ GeV}, \quad
  972.319\mbox{ GeV}, \quad
  2140.13\mbox{ GeV} \\ 
  & {\rm Set\ 3:} & \quad \quad 43.6216\mbox{ GeV}, \quad
  982.737\mbox{ GeV}, \quad
  4114.49\mbox{ GeV} ~,
  \end{eqnarray}
and those of matrix $M^2_{(1,3]}$ are found to be 
\begin{eqnarray}
  & {\rm Set\ 1:}& \quad \quad 74.7636\mbox{ GeV}, \quad
  1675.68\mbox{ GeV}\\
  & {\rm Set\  2:}& \quad \quad 44.8651\mbox{ GeV}, \quad
  1008.779\mbox{ GeV} \\
  & {\rm Set\ 3:}& \quad \quad 43.9536\mbox{ GeV}, \quad
  987.883\mbox{ GeV}.
\end{eqnarray}
The $[G_{\rm NC}(Q^2)]_{WW}$  results for Sets 1, 2, and 3 are depicted in the upper, middle, and
lower curves of Figure~\ref{fig:WWtex}.
Set 1 is, again,  indistinguishable from the standard
model.
For the Set 2 parameters, the deviation of the correlation function from its standard model form is less than 0.5\% even at $\sqrt{-Q^2}=200$ GeV; this choice of parameters seems to be phenomenologically acceptable.  For Set 3, the deviation is about 2\% at $\sqrt{-Q^2}=200$ GeV, which is too
large to be phenomenologically acceptable. 
  
\subsection{Electroweak corrections}

By construction, we expect that the electroweak corrections other than
$\alpha\delta$ will be suppressed for any of our sample sets of parameters.  Because the 
model is Case I \cite{SekharChivukula:2004mu}, $\Delta\rho = 0$; because 
$g_{N+1}$ is small (\ref{eq:smallgs}) , $\alpha T \approx 0$; the value of 
$x_1$ was explicitly chosen (\ref{eq:newlimiti}) to make $\alpha S \approx 0$.  This turns out to 
be the case; numerical evaluation shows
$  |\alpha S| \lessim 10^{-5}, \, 
  |\alpha T| \lessim 10^{-5}, \, 
  |\alpha U| \lessim 10^{-5}$.

However, the value of $\alpha\delta$ is not automatically small enough to agree with constraints set by data.  Set 1 has the smallest value of $x_1$, and the corresponding value of $\alpha \delta$ is
zero to within the limits of numerical accuracy.
For Set 2, the experimental upper bound \cite{Barbieri:2004qk} of order .001 is satisfied by the quantity 
\begin{equation}
  \dfrac{\alpha\delta}{4s_Z^2 c_Z^2} = 1 - \dfrac{[\xi_W]_{WW}}{4\sqrt{2}G_F M_W^2}
  = 0.70 \times 10^{-3},  
\end{equation}
Note that the approximate value for $\alpha\delta$ from equation (\ref{eq:alphadelta}) is consistent with the exact result above: the difference is precisely the size of the terms neglected in the approximation.
For set 3, on the other hand, $\alpha\delta$ lies above the experimental bound
\begin{equation}
  \dfrac{\alpha\delta}{4s_Z^2 c_Z^2}
  = 3.04\times 10^{-3}.
\end{equation}
In other words, choosing the amount of delocalization to guarantee that the oblique correction $\alpha S$ is small does not guarantee that the universal non-oblique correction $\alpha\delta$ will be of acceptable size.  The first requires  $x_1$ to be a function of the couplings and $f$-constants; the second places an absolute upper bound on the value of $x_1$.  In our 4-site model, the most significant effect of the larger value of $f_1$ for set 3 was to drive $x_1$ larger -- which pushed $\alpha\delta$ too high.

\section{Conclusions}

In this note we have calculated the form of electroweak corrections
in deconstructed Higgsless models
for the case of a fermion whose weak properties arise from two adjacent $SU(2)$
groups on the deconstructed lattice. 
We have shown that, as recently proposed in the continuum \cite{Cacciapaglia:2004rb,Foadi:2004ps}, 
it is possible for the value
of the electroweak parameter $\alpha S$ to be  small in such  a model. 
Working in the deconstructed
limit we have also directly evaluated the size of $\alpha \delta$, arising off-$Z$-pole 
from the exchange of Kaluza-Klein modes \cite{Chivukula:2004af}.  This has not previously been evaluated in the continuum. 
In one-site delocalized Higgsless models with small values of
$\alpha S$, we showed that the amount of
delocalization is bounded to be less than of order 25\% at 95\%CL due to the simultaneous need to ensure unitarization of 
$W_LW_L$ scattering and to provide a value of $\alpha\delta$ that agrees with experiment.
We have discussed the relation of these calculations to our previous calculations in
deconstructed Higgsless models \cite{SekharChivukula:2004mu}, 
and to models of extended technicolor. Finally, 
we presented numerical results for a four-site model, illustrating our
analytic calculations.
In a subsequent publication, \cite{newwork}, we will generalize our discussion to
multi-site delocalization and discuss the effects of fermion delocalization in the continuum.

\acknowledgments

R.S.C. and E.H.S. are supported in part by the US National Science Foundation under
award PHY-0354226.
M.K. acknowledges support by the 21st Century COE Program of Nagoya University 
provided by JSPS (15COEG01). M.T.'s work is supported in part by the JSPS Grant-in-Aid for Scientific Research No.16540226. H.J.H. is supported by the US Department of Energy grant
DE-FG03-93ER40757.



\end{document}